# VERIFICATION OF AN AGENT-BASED DISEASE MODEL OF HUMAN MYCOBACTERIUM TUBERCULOSIS INFECTION

Cristina Curreli[1,2], Francesco Pappalardo[3], Giulia Russo[3,4], Marzio Pennisi[5], Dimitrios Kiagias[6], Miguel Juarez[6], Marco Viceconti[1,2]

[1] Department of Industrial Engineering, Alma Mater Studiorum - University of Bologna (IT).

[2] Medical Technology Lab, IRCCS Istituto Ortopedico Rizzoli, Bologna (IT).

[3] Department of Drug Sciences, University of Catania (IT).

[4] Mimesis srl, Catania (IT).

[5] Computer Science Institute, DiSIT, University of Eastern Piedmont, Alessandria, Italy

[6] School of Mathematics & Statistics and Insigneo and Institute for in silico Medicine, University of Sheffield (UK).

CORRESPONDING AUTHOR:

Dr Cristina Curreli

Medical Technology Lab

Istituto Ortopedico Rizzoli

Via di Barbiano 1/10, 40136 Bologna (IT)

Email: cristina.curreli@unibo.it

# VERIFICATION OF AN AGENT-BASED DISEASE MODEL OF HUMAN MYCOBACTERIUM TUBERCULOSIS INFECTION

**ABSTRACT**

Agent-Based Models are a powerful class of computational models widely used to simulate complex phenomena in many different application areas. However, one of the most critical aspects, poorly investigated in the literature, regards an important step of the model credibility assessment: solution verification. This study overcomes this limitation by proposing a general verification framework for Agent-Based Models that aims at evaluating the numerical errors associated with the model. A step-by-step procedure, which consists of two main verification studies (deterministic and stochastic model verification), is described in detail and applied to a specific mission critical scenario: the quantification of the numerical approximation error for UISS-TB, an ABM of the human immune system developed to predict the progression of pulmonary tuberculosis. Results provide indications on the possibility to use the proposed model verification workflow to systematically identify and quantify numerical approximation errors associated with UISS-TB and, in general, with any other ABMs.

## 1. INTRODUCTION

The evolution of computer models of human pathophysiology enables predicting the response of individual patients to various treatment options (Digital Patient solutions) [1] or to predict the efficacy and/or safety of new treatments on a cohort of virtual patients (In Silico Trials solutions) [2]–[4]. But before such technologies can be used in these mission-critical scenarios, their *credibility* must be thoroughly assessed [5]. Emerging technical standards [6] and recent regulatory guidelines [7] propose general rules and frameworks that can be used to assess model credibility through *verification*, *validation*, and *uncertainty quantification* (VV&UQ) studies. Although the general aims of the different credibility activities are well established and valid for all types of models, the detailed definition of the tests to be performed may vary depending on the computational approach used. This is particularly true for model verification, a part of the verification process that aims to identify, quantify and reduce the numerical error associated with the model [8].

In this paper, we focus on the part of model verification called solution or calculation verification, of Agent-Based Models (ABMs), a topic that is still considered one of the main challenges in the field [9]. ABMs are a powerful class of computational models widely used in many different application areas including social science [10], ecology [11] and biology [12]. The increasing interest in ABMs in the biomedical field is mainly due to their ability to simulate complex phenomena adopting a bottom-up approach: the global behaviour of the heterogeneous real system is modelled using autonomous entities of different nature called agents that can interact and evolve based on their internal state and on different environmental factors. Depending on how their interaction and state-transition rules are defined, ABMs may be seen as mechanistic models; however, compared to conventional equation-based mechanistic models, the overall system behaviour is not known a priori and the interaction of the discrete autonomous agents is mainly described by cause-effect relationships. Another characteristic of ABMs worth considering when dealing with model verification, is their stochastic nature: the interaction of agents is often defined using deterministic approaches (e.g., molecular dynamics) but the intrinsic randomness of the complex phenomena modelled (e.g., spatial distribution of agents) is introduced using stochastic variables [13].

In models where the underlying theory is explicitly expressed in term of mathematical equations, the numerical error we commit in solving these equations can be due to how we implemented in software the solution process (*code verification*), or by the assumptions we make to obtain a numerically approximated solution, such as space-time discretisation (*solution verification*). But in ABM the underlying theory is expressed in term of local rules; this makes it somehow challenging to define exactly what model verification should be. The difficulty of establishing verification methods for ABMs is mainly due to these fundamental features [9], [14] and only few studies try to meet this challenge. Ormerod and Rosewell [15] highlight the fact that widely recognized verification approaches such as Runge-Kutta methods for numerical solution of differential equations cannot be used for ABMs. They consider replicability, distributional equivalence and relational alignment key aspects of verification. The first two concepts are part of verification: replicability means verifying that the model exactly produces the same results if the same inputs are used; distributional equivalence is the identification of the number of times the stochastic model has to be solved to establish its properties. But relational alignment, which involves comparing the model predictions with expected trends, is instead part of validation. In their ABMs verification study, Guoyin et al. [16] first tested that the logic structure of the simulation codes was in accordance with the theoretical model, and then conducted software verification tests such as unit, integration and case tests; in essence, reducing verification to code verification, and leaving out model verification. A similar limited approach is reported also in [17]. Leombruni et al [14] discuss the need of a common protocol for agent-based simulations and present a detailed description of the analyses that should be conducted once the model is defined. However, the distinction between model verification and validation is not very clear and typical code verification activities like bug tracking are considered part of the validation process. In [18], a generic software testing framework for ABMs, mainly focused on validation in the sense of software construct, has been proposed and applied on a realistic simulation case study.

To the extent of the authors knowledge, there are no studies in the literature that describe a step-by-step procedure for solution verification of ABMs. Defining a rigorous verification framework that investigates all these points is of fundamental importance, especially when dealing with mission-critical applications.

The aim of this work is to propose a detailed solution verification scheme for ABMs for In Silico Trials. As exemplary case to demonstrate the proposed approach, we use UISS-TB, an ABM of the human immune system [19] developed to predict the progression of pulmonary tuberculosis.

## 2. MATERIALS AND METHODS

### 2.1. UISS-TB

*Main modelling features*

The Universal Immune System Simulator (UISS) is an agent-based model of the human immune system, which accounts for both innate and acquired immune response; UISS has been used effectively to model the response of the immune system to a variety of diseases [20] [21]. Within the StriTuVaD project UISS has been extended to model the response of the immune system to the pulmonary infection of Mycobacterium Tuberculosis (MTB). In a preliminary study, it has been shown that the resulting simulator (UISS-TB) could be used to test *in silico* the efficacy of new therapies against tuberculosis, one of the most deadly infectious diseases in the world [22].

The process of MTB infection is modelled in UISS-TB as a series of interactions between autonomous biological entities such as pathogens, cells or molecular species. The most important interactions between these agents can be described as a function of their type and state, their vicinity, the concentration of certain chemical species in the neighbourhood, and/or the molecular fingerprint (presentation patterns) that each entity exposes. The anatomical compartment of interest is modelled with a cartesian lattice mathematical structure [23], a bidimensional domain whose state is dynamic in time. Within this spatial domain, of appropriate size and specific to the phenomena modelled, cells can differentiate, replicate, become active or inactive, or die. Pathogens and other molecular entities can replicate or die. An important modelling feature regards the receptor-ligand affinity and specificity: all immune repertories, also called cell receptors, are represented as a set of binary strings, and binding events are modelled with bit string matching rules mainly based on the complementary Hamming distance [24].

*Deterministic and Stochastic nature of UISS-TB*

Conceptually, all interactions in UISS-TB are stochastic in nature. But from an implementation point of view, this is achieved through *random seeds* (RSs) produced by pseudo-random generators. This means that if the same value of random seed is used the interaction rule will produce the same result, deterministically. This makes possible to separate, for the purpose of solution verification, the deterministic and stochastic aspects of the model, and investigate them separately.

UISS-TB uses three different RSs representing three stochastic variables: initial distribution of the agents ($RS_{id}$), randomization of the environmental factors (e.g., the effect on the lymphatic flow and cell density constraints; ($RS_{ef}$)), and the human leukocyte antigen type I and II (HLA-I and HLA-II; ($RS_{HLA}$)). Different pseudo number generator algorithms (MT19937 [25], TAUS 2 [26] and RANLUX [27], respectively) are used for the RS initializations.

*Inputs/ Outputs model definition*

The UISS-TB model is informed by a set of $N_I$ = 22 inputs, hereinafter named *vector of features*, comprising measurable quantities in an individual MTB patient. Among these are significant components of the adaptive immune system (e.g., lymphocytes cells, cytokines immunoregulatory molecules, interferons), main characteristics of the patient (e.g., age and body mass index) and typical MTB disease features such as initial bacterial load in the sputum smear and bacterial virulence. Table 1 lists all 22 inputs, with their admissible minimum and maximum values. Data reported in [28], [29] are used as a reference to define the main characteristic of the patients.

To provide an exemple of the verification approach proposed, we used in this study the immune system response simulator of a healthy individual that at time 0 is exposed to an infective challenge by MTB. Usually, when the human immune system faces an antigen for the first time, its recognition process with the innate immune arm is triggered. Later, the adaptive response is mounted specifically for that antigen. If at a later stage the same antigen is recognized by the same host immune system, a stronger and faster immune response starts directly with the adaptive arm: this can be recognised by observing a higher peak in both specific immunoglobulin and cellular responses directed against the pathogen. The input set used to represent the so called "Newly Infected Patient" (NIP) are also reported in Table 1. It should be stressed that this input set does not represent any specific individual, but a possible NIP.

*Table 1. UISS-TB vector of features definition: admissible minimum ($I_{MIN}$), maximum ($I_{MAX}$) values and input settings for the NIP case study ($I_{NIP}$).*

| INPUTS | DESCRIPTION | $I_{MIN}$ | $I_{MAX}$ | $I_{NIP}$ |
|---|---|---|---|---|
| **MTB_VIR** | Virulence Factor | 0 | 1 | 0.5 |
| **MTB_SPUTUM (CFU/mL)** | Bacterial Load in the sputum smear | 0 | 10000 | 0 |
| **TH1 (CELLS/µL)** | CD4 T cell type 1 | 0 | 100 | 0 |
| **TH2 (CELLS/µL)** | CD4 T cell type 2 | 0 | 100 | 0 |
| **IgG (TITER)** | Specific antibody titer | 0 | 512 | 0 |
| **TC (CELLS/µL)** | CD8 T cell | 0 | 1134 | 0 |
| **IL-1 (pg/mL)** | Interleukin 1 | 0 | 235 | 0 |
| **IL-2 (pg/mL)** | Interleukin 2 | 0 | 894 | 0 |
| **IL-10 (pg/mL)** | Interleukin 10 | 0 | 516 | 0 |
| **IL-12 (pg/mL)** | Interleukin 12 | 0 | 495 | 0 |
| **IL17-A (pg/mL)** | Interleukin 17A | 0 | 704 | 0 |
| **IL-23 (pg/mL)** | Interleukin 23 | 0 | 800 | 0 |
| **IFN1A (pg/mL)** | Interferon alpha-1 | 0 | 148.4 | 0 |
| **IFN1B (pg /mL)** | Interferon beta-1b | 0 | 206 | 0 |
| **IFNG (pg/mL)** | Interferon gamma (IFNγ) | 0 | 49.4 | 0 |
| **TNF (pg/mL)** | Tumor Necrosis Factor | 0 | 268.2 | 0 |
| **LXA4 (ng/mL)** | Lipoxin A4 | 0 | 3 | 0 |
| **PGE2 (ng /mL)** | Prostaglandin E2 | 0 | 2.1 | 0 |
| **VITAMIND (ng/mL)** | Vitamin D | 25 | 80 | 25.8 |
| **TREG (CELLS /µL)** | Regulatory T cells | 0 | 200 | 55 |
| **AGE (YEARS)** | Age | 10 | 80 | 35 |
| **BMI (Kg/m$^2$)** | Body Mass Index | 18 | 35 | 31 |

In order to monitor the disease progression, four chemical species were selected as output variables (OVs): T helper 17 cells (Th17), Antibodies (Ab), B cells (B) and Antigens (TotAGS). Th17 lymphocytes represent one of the main subsets of helper T cells, primarily involved in recruiting neutrophils. Ab are circulating proteins produced in response to the exposure to antigens. B lymphocytes are a type of white blood cell which principal role is to secrete antibodies. An antigen is a molecule composed by the epitopes and the peptides of the pathogen that can bind to an antigen-specific antibody or B cell antigen receptor.

For all these entities, UISS-TB traces the evolution of their concentration (number of entities in the entire space domain that represents 1 µL of peripheral blood sample) through the simulation time (one year). The concentration curves were characterized by three key features:

- *Peak Value* (PV): maximum value of the time series in correspondence of the second infection peak;

- *Time to Peak Value* (TPV): time at which the PV occurs;
- *Final Value* (FV): value of the time series at the end of the simulation.

Hence, we analyse 12 output quantities, Mo: three key features (PV, TPV and FV), for each one of the four output variables of interest Th17, Ab, B and TotAGS.

*2.2. General model verification workflow*

As already mentioned, ABMs are not formulated in terms of a set of related equations (such as ordinary differential equations), but rather by a set of interaction rules and state transitions. Thus, in addition to the verifications that are typical of mathematical models (e.g. quantification of the discretisation error) we also need an alternative for analytical checks of properties from solutions to sets of related equations (e.g. existence, uniqueness, smoothness, non-chaoticity).

The general model verification workflow adopted in this study consists in the sequence of steps depicted in Figure 1 and described in detail in this section. Two main verification studies can be identified: deterministic and stochastic model verification. The first includes four different analyses that aim at (i) verifying basic essential properties such as existence and uniqueness of the solution, (ii) quantifying the discretization error associated with solving the computational problem at a finite number of temporal grid points, (iii) evaluating the smoothness of the output time series data, and (iv) exploring critical parameter settings and their corresponding output measures. All these deterministic verification tests were performed considering fixed values of the RSs. While $RS_{id}$ and $RS_{ef}$ are set randomly, $RS_{HLA}$ is chosen to represent the mean HLA-I and HLA-II that belong to Caucasian population. Since stochasticity is part of UISS-TB, the second verification study aims at investigating the effect of the randomization factors. In order to assess statistical robustness, the model is run multiple times varying the RSs and the distribution of outcomes are studied in terms of consistency and variance stability. The determination of the minimum sample size to reach a good statistical significance is also investigated. For simplicity, the probabilistic verification analyses were performed in this study only considering the effect of randomising the environmental factors. The random seed that is responsible of defining this stochastic process, $RS_{ef}$, is thus varied in order to produce a number of S = 1000 different simulations. The other model inputs are set according to $I_{NIP}$ (Table 1).

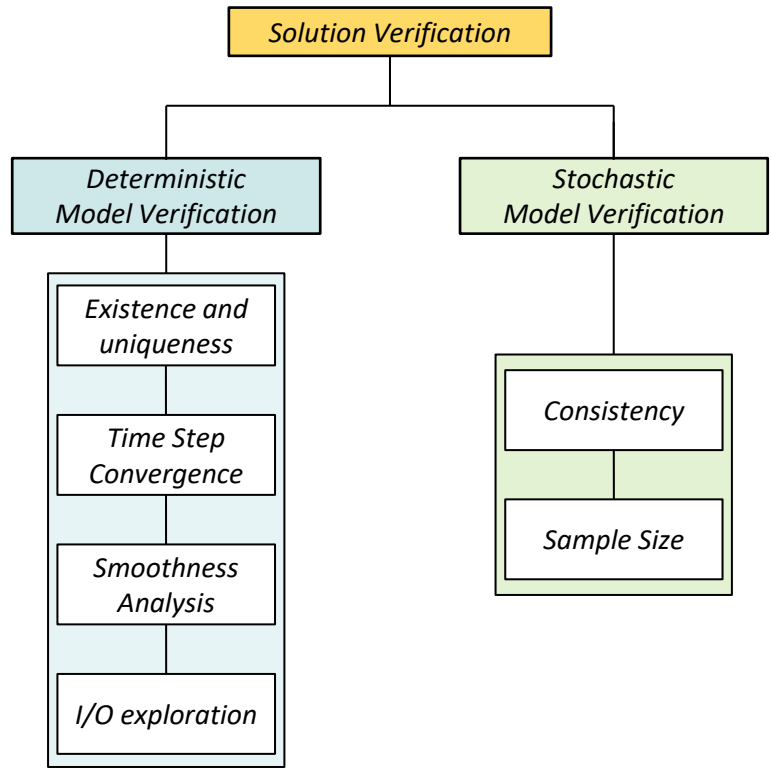

*Figure 1. General solution verification workflow which consists in two main part: deterministic and stochastic model verification.*

*Existence and uniqueness*

The first fundamental model verification test regards existence and uniqueness of the solution. Because time is considered in discrete steps, all the theorems for continuously differentiable equations cannot be applied directly. Existence of the solution was checked by verifying that for all the feature sets within the admissible range, UISS-TB returned a valid output set; uniqueness was checked by ensuring that identical input sets would always produce the same output results.

*Time step convergence analysis*

The only discretisation error in UISS-TB is the one associated with solving the computational problem at a finite number of temporal points called time steps (*TS*). All molecules and cells handled by UISS-TB are considered as single entities. To evaluate the effect of the time discretization on the model predictions, the output time series data representing the disease progression for each of the OVs were compared by setting different *TS* sizes. We explore a range of time steps between 48 hours and one minute. Two days was considered the temporal resolution that was required clinically, whereas the one minute value was set on the basis of an acceptable computational cost.

All the other inputs were kept constant, including the RSs. Convergence was checked for all the OVs on the three different quantities $q$ = PV, TPV and FV, by computing the percentage discretization error according to the relationship:

$$e_q^i = \frac{q^{i*} - q^i}{q^{i*}} \cdot 100 \qquad (1)$$

where superscript $i=1..13$, refers to the time series obtained by setting the time step size $TS^i$ according to Table 2 and characterized by a number of iterations $N^i$. The results obtained with the model $i^* = 14$ (*TS* is set to its minimum value of 1 minute) were considered as reference solutions for the analysis. If $q^{i*}$ was equal to zero, the discretization error was computed using the mean value of the time series obtained with model $i^*$ as denominator of Equation (1). The model was assumed to converge if the error $e_q^i < 5\%$. Additionally, the correlation coefficient was computed to investigate global similarity between curves.

*Table 2. Temporal grid point refinement scheme used for the time step convergence check*

| i | $TS^i$ (min) | $N^i$ |
|---|---|---|
| 1 | 2,880 | 182 |
| 2 | 1,440 | 365 |
| 3 | 960 | 547 |
| 4 | 480 | 1,095 |
| 5 | 240 | 2,190 |
| 6 | 120 | 4,380 |
| 7 | 60 | 8,760 |
| 8 | 30 | 17,520 |
| 9 | 20 | 26,280 |
| 0 | 15 | 35,040 |
| 11 | 10 | 52,560 |
| 12 | 5 | 105,120 |
| 13 | 2 | 262,800 |
| 14 | 1 | 525,600 |

*Smoothness analysis*

As mentioned in the introduction, calculation verification studies aim at assessing the errors that may occur in the numerical solution. Among those there are singularities, discontinuities and buckling. One metric that can be used to evaluate the presence of such problems is smoothness.

In order to evaluate the smoothness of the solution, a coefficient of variation *D* was computed for all output time series (evolution of the OVs chemical species concentration) obtained with the UISS-TB simulator. The value of the TS was set according to the results obtained with the time step convergence analysis presented in Section 3.1. The coefficient *D* represents the standard deviation of the 1$^{st}$ difference of the time series scaled by the absolute value of their mean and is a measure of how variable the discrete first derivative of the time series is along the whole time series: the smaller D is the smoother the series. In order to mimic a regular smoothness analysis and detect possible local roughness in the time series, D has been computed using moving windows. For each observation in the time series, $y_t$, the *k* nearest neighbours are considered in the windows: $y^k_t = \{y_{t-k}, y_{t-k+1}, \ldots, y_t, y_{t+1}, \ldots, y_{t+k}\}$. All the measurements were applied in this study with moving windows of size *k* =3.

*Parameter sweep analysis*

The goal of the parameter sweep analysis is to ensure that the model is not numerically ill-conditioned. The entire input sets space is sampled uniformly looking for particular input sets for which the model fails to produce a valid solution, or the solution provided is outside the range of expected values. Last, the analysis checks whether small variations of the input values produce large variations of the output values, which would suggest an abnormal sensitivity to some inputs. Since are frequently extreme input values that cause problems, a Design of Experiment (DoE) approach is preferred to a Monte Carlo sampling, so to ensure that all extreme input values are tested. But in this case, the high dimensionality of the input space (22 inputs), would make a DoE computationally prohibitive. Thus, a hybrid approach was adopted. An initial sensitivity analysis was carried out by perturbing each input once at a time from their minimum to their maximum value, while holding the rest at their mean values. A "reference input vector of features" was thus first defined using as input setup the mean values for each of the $N_I$ inputs (Table 1). Considering all the $M_o$ outputs (PV, TPV, FV for each of the four OVs) obtained with all the 44 (2x$N_I$) simulations, a matrix of $N_I$ x $M_o$ coefficients $p_{v,j}$ was computed to determining the inputs that mostly affect the outputs:

$$p_{v,j} = \frac{(o_j^{v_{max}} - o_j^{v_{min}})/o_{j_{med}}}{(I_{v_{max}} - I_{v_{min}})/I_{v_{med}}} \cdot 100 \qquad \text{with } v = 1,\ldots, N_I \text{; and } j = 1,.., M_o \qquad (2)$$

where $O_j^{v_{max}}$ and $O_j^{v_{min}}$ are the values of the output j obtained by setting the input *v* respectively to its maximum and minimum value; $O_{j_{med}}$ is the output *j* obtained with the reference input vector of features; $I_{v_{max}}$, $I_{v_{min}}$ and $I_{v_{med}}$ are respectively the maximum, minimum and mean value of the model input *v*.

Latin hypercube sampling (LHS) was then used to generate sample of P = 100 values on each of the 10 input variables that mostly affect the output results. The global variation effect on each output *j* was quantified by computing a coefficient of variation $C_j$:

$$C_j = \frac{\sigma_j}{\mu_j} \tag{3}$$

where $m_j$ and $s_j$ are respectively the mean and standard deviation from output *j*.

*Consistency*

The outputs obtained from the S =1000 simulations varying the $RS_{ef}$ factor, were studied in terms of statistical consistency. The *TS* size was selected for all simulation based on the time step convergence analysis results (Section 3.1).

The shape of the sample distributions for all the Mo outputs was first checked by investigating their fit to Gaussian and Student-t distributions using Kullback-Leibler (KL) divergence measure [30]: KL divergence is non-negative and zero if and only if both distributions are exactly the same (i.e., the closer the KL to nought, the more similar the distributions). Because the scale of KL is arbitrary, the ratio *r(X)* of the divergence of the Student-t ($KL_s$) and the Gaussian ($KL_G$) to the distribution of the data, was computed to identify the "best" fitting shape with the following relationship:

$$r(X_j) = \frac{KL_S(X_j)}{KL_G(X_j)} \tag{4}$$

where $X_j$ is the distribution of the natural logarithm of the output *j*.

Mean, standard deviation, median, 25th, and 75th percentile are also used to characterize the distributions.

*Sample size determination*

The minimum sample size, *Ns*, to have a sufficiently accurate estimates of the average and standard deviations are calculated, for each output *j*, based on the following criterion:

$$IC_s = \left| \frac{CoV_{s+1} - CoV_s}{CoV_s} \right| < \varepsilon \tag{5}$$

where the subscript *s* refers to the number of samples (i.e., simulations run) obtained with a different $RS_{ef}$ value, and the coefficient of variation index ($CoV_s$) is defined as the ratio of the cumulative standard deviation (CStd_s) and mean (CMean_s) for increasing sample size *s*.

In this study, convergence and stability of the mean and standard deviation was assumed to be satisfied for $\varepsilon = 0.01$. The minimum sample size *Ns* was thus defined as the value of *s* such that $IC_s$ is less than 0.01. It is worth noting that the value of $\varepsilon$ is arbitrary and might be selected according to the level of accuracy required.

## 3. RESULTS

### 3.1. Time step convergence analysis

As already stressed, the quantification of the discretization error is a fundamental step in model verification. In this case, the convergence analysis allows the identification of a sufficient TS value balancing a compromise between two opposing requirements: a large value of TS to reduce the computational cost and a small value of TS to guarantee that the discretization error does not exceed for each predicted quantity a value of 5%.

For this case study, the quantities that show fast convergence as the time step decreases are TPV and FV (Fig. 2(a,b)): for all the OVs, the discretization error associated with these outputs reached a value less than 5% at TS= 10 minutes that correspond to 52,560 computer iterations and a simulation time of about 8 minutes. The values at 12 months were always equal to zero for Ab and TotAGS, regardless of the time step, while FV of Th17 and B converged monotonically and asymptotically to 0 at a similar rate (Fig. 2(b)). Percentage differences *e* (%) of about 2% and 1% were observed at TS= 10 min for the FV of Th1 and B, respectively (Fig. 2(b)). As far as TPV is concerned, for all the OVs any TS of 16 hours or smaller produced percentage differences less than 5% (Fig. 2(a)).

Peak value showed convergence only for TotAGS: the percentage difference rapidly reached a value less than 5% (3.1%) for *i* = 2 that correspond to a TS of 24 hours. A severely oscillatory behaviour was instead observed especially for B, Ab and Th17 with no signs of convergence (Fig. 2(c))

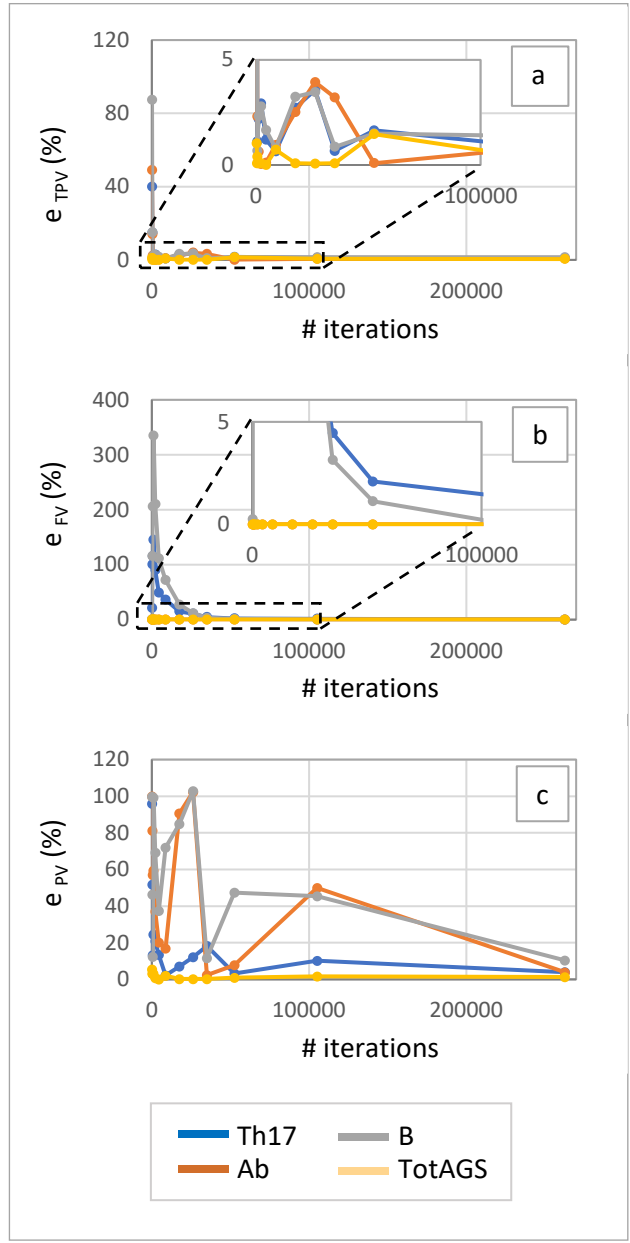

*Fig 2. Trend on the percentage approximation error $e_{Th17}$, $e_{Ab}$, $e_B$ and $e_{TotAGS}$ with decreasing values of the TS for the three quantities TPV (a), FV (b) and PV (c).*

### 3.2. Smoothness analysis

Here, local roughness in the time series is assessed. Although the overall trend can be considered smooth for all the OVs with values of the local coefficient *D* close to zero (Fig. 3), in the regions around the two peaks was possible to observe higher values of the standard deviation of the first discrete derivative. This is evident for Th17 and especially for Ab where *D* reaches a maximum value of about 50 and 890 respectively (Fig. 3).

Maximum values of the first discrete derivative, relative to the maximum chemical species concentration rate, were about 4.3, 73.7, 4 and 166.7 (# entities/ (µL *s)) for Th17, Ab, B and TotAGS, respectively.

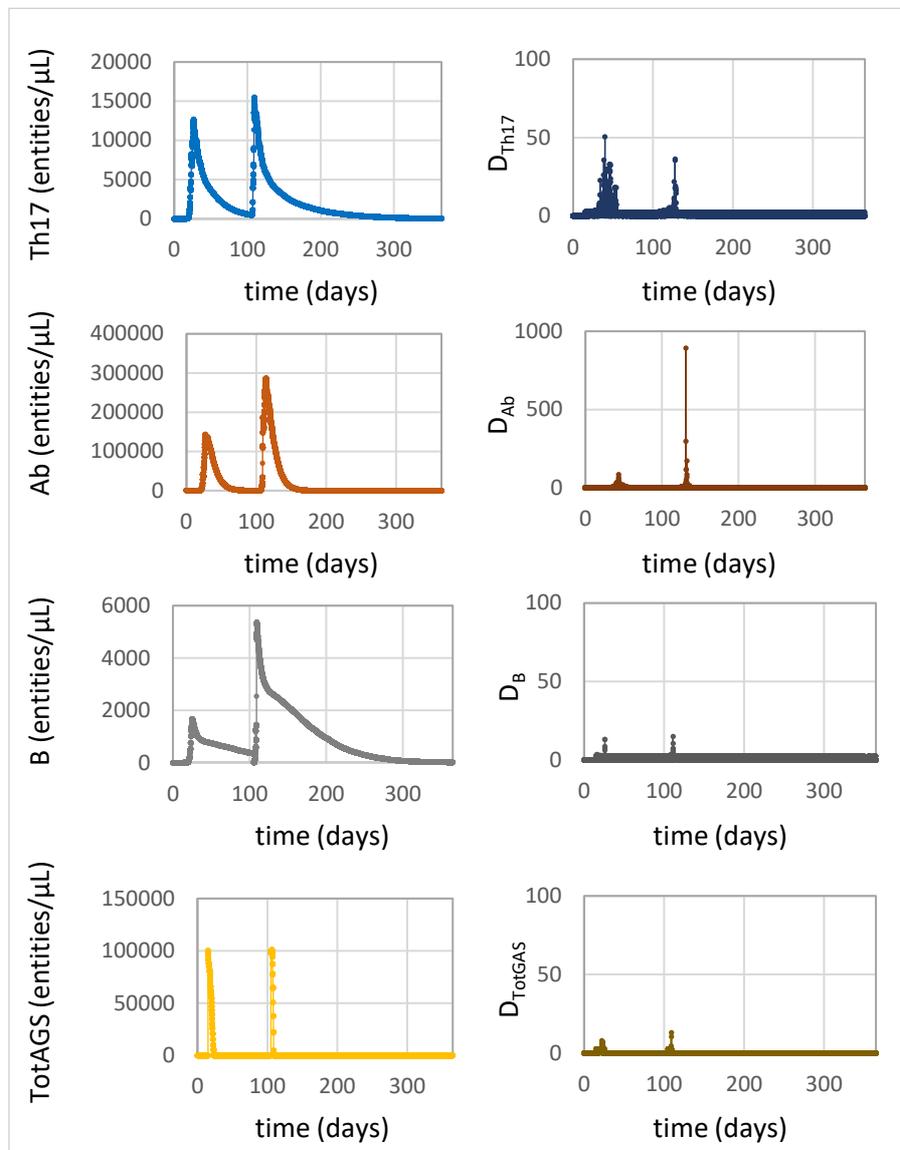

*Fig 3. Evolution of the OVs concentration during the simulation time of one year (left) and relative smoothness measures (right).*

### 3.3. Parameter sweep analysis

The initial "one at the time" sensitivity analysis performed enables identification of the inputs with a higher effect on the outputs based on the *p* coefficient defined in Section 2.2. These are: Mtb_Vir, BMI, AGE, TC, IgG, PGE2, TNF, IL-12, Th1 and Th2. The most sensitive responses are the final value of Th17 and the peak value of Ab due to changes of the BMI input ($p_{BMI,FV\_Th17} = 272\%$, $p_{BMI,PV\_Ab} =$ 149%). From this initial I/O exploration it is also possible to notice that all the $M_o$ outputs appeared to be completely insensitive to 2 inputs: Treg and LXA4, while the outputs whose values do not change with the inputs are FVs of Ab and TotAGS.

A summary of the results obtained from the second part of the parameter sweep analysis in terms of coefficient of variation $C$ is reported in Fig. 4. The largest variation observed was for the FV of Th17 and B ($C_{FV\_Th17}$ = 64.7% and $C_{FV\_B}$ = 59.3%); for all the other outputs, $C$ was always below 35%. It is worth noting that FVs of Ab and TotAGS did not change with the P = 100 different input combinations tested, and the coefficient $C$ was thus 0.

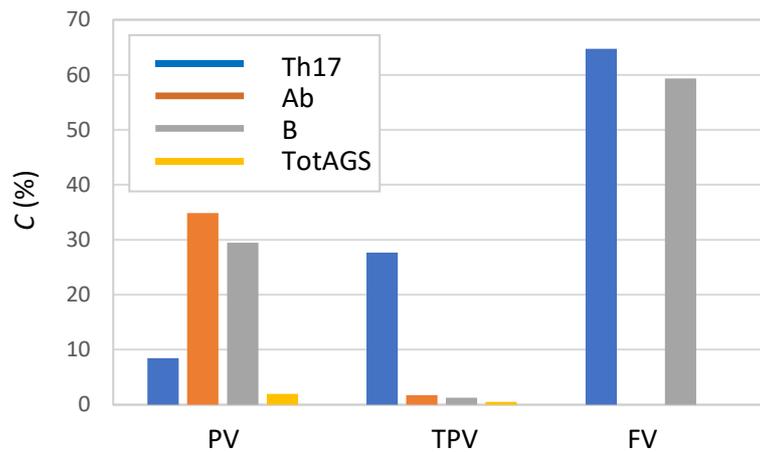

*Fig 4. Coefficient of variation C for all the output variables*

### 3.4. Consistency

Details of the results obtained with the statistical analyses in term of mean, standard deviation, median, 75 and 25 percentiles are reported in Table 3. In all cases, the mean and median have similar values and the coefficient of variation (not shown) is below 0.33. The ratio $r(X)$ between the KL divergence of the Student-t and the Gaussian fit to the distribution of the data is also reported in Table 3.

### 3.5. Sample size determination

The required minimum sample size $Ns$ for all the Mo output variables are reported in Table 3. For the sake of completeness, an example of the trend of the cumulative mean, standard deviation and $IC$ index with increasing number of $s$ is shown in Fig. 5. It is possible to notice that, for the output variable $PV_{Ab}$, an accurate estimate of the mean and standard deviation is achieved when the simulation is run 293 times varying the $RS_{ef}$ factor.

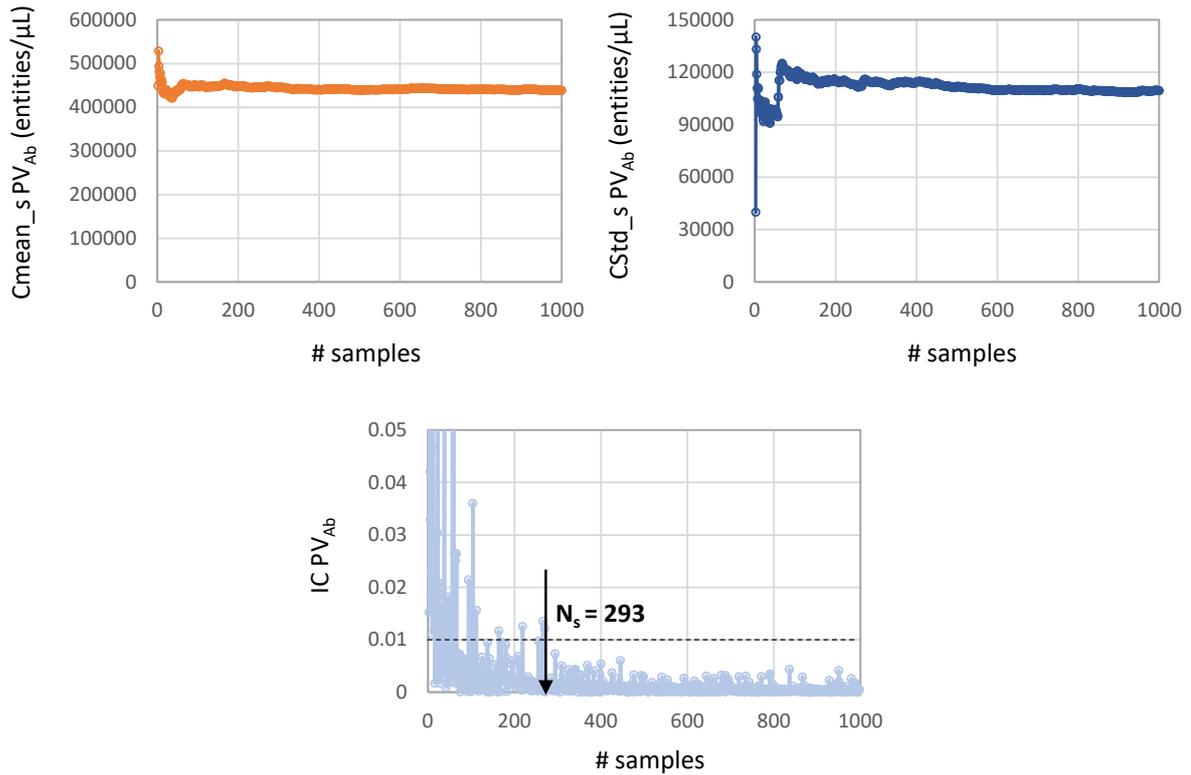

*Fig 5. Trend of the CMean_s (a), CStd_s (b) and IC (c) index with increasing values of s for the output variable $PV_{Ab}$ and identification of the minimum sample size Ns required to have an accurate estimate of the mean and standard deviation.*

*Table 3. Statistic summary for all the output distributions in term of mean, standard deviation, median, 75 and 25 percentiles. Quantities related to PV and FV are expressed in # entities/μL while those related to TPV are expressed in hours. Non dimensional measures r(X) and Ns are also reported.*

| Output | | Mean | St. Deviation | Median | Pctl (75) | Pctl (25) | r(X) | Ns |
|---|---|---|---|---|---|---|---|---|
| **Th17** | PV | 15529.892 | 988.233 | 15574 | 14869 | 16153 | 1.064 | 683 |
| | TPV | 2447.792 | 583.817 | 2626.083 | 2604.75 | 2649.833 | 0.515 | 594 |
| | FV | 69.338 | 17.721 | 69 | 58 | 80 | 0.835 | 497 |
| **Ab** | PV | 439569.193 | 109589.85 | 429671.5 | 362208 | 506448 | 1.148 | 293 |
| | TPV | 2727.7 | 45.627 | 2727.33 | 2695.5 | 2757.16 | 1.085 | 440 |
| | FV | 0 | 0 | 0 | 0 | 0 | - | - |
| **B** | PV | 6142.913 | 792.342 | 6158.5 | 5597.5 | 6698.5 | 1.065 | 497 |
| | TPV | 2618.283 | 27.478 | 2615.166 | 2598.5 | 2634.83 | 1.131 | 249 |
| | FV | 17.459 | 5.72 | 17 | 13 | 21 | 1.188 | 340 |
| **TotAGS** | PV | 100786.779 | 900.267 | 100573 | 100235 | 101106.5 | 0.836 | 851 |
| | TPV | 2528.387 | 9.536 | 2525.5 | 2522.16 | 2530.75 | 0.831 | 917 |
| | FV | 0 | 0 | 0 | 0 | 0 | - | - |

## 4. DISCUSSION

The aim of this study was to develop a detailed solution verification procedure that can be used to identify and quantify numerical errors associated with a generic ABM. A step-by-step approach mainly based on a verification strategy that separates the deterministic and stochastic solution verification was proposed. The deterministic solution verification did include aspects commonly present in the verification of models based on differential equations, such as the effect of time stepping, and aspects that are typically part of the verification of data-based models, such as existence and unicity, smoothness, etc. The validity of the ABM verification procedure was tested by applying all the verification activities to a specific case study: UISS-TB, an agent-based model that can be used to test *in silico* the efficacy of new therapies against tuberculosis.

### *4.1. Deterministic model verification*

From the first deterministic model verification activities, performed with fixed values of all the stochastic variables, was possible to:

   i.   ensure that, for all the different input feature set, UISS-TB estimates an acceptable and unique output set;

   ii.  detect uncertainties associated with solving the computational problem at a finite number of temporal grid points;

   iii. identify local roughness in the output time series data that might indicate unrealistic increment rate of the output quantity of interest;

   iv.  determine the degree to which inputs parameter values affect the outputs and verify that any combinations of inputs do not lead to unexpected trends of the model's results.

The results obtained from the time step convergence analyses, presented in Section 3.1, suggest that a TS of 10 minutes is enough to accurately predict both TPV and FV. The discretization error was in fact always less than 5% for all the four chemical species of interest. However, except for $e_{TotAGS}$, the trend of the percentage approximation error for PV showed an oscillatory behaviour. The concentration of B cell output predicted with a time step of 2 mins was about 10% higher than that predicted with a time step of 1 minute (minimum acceptable value of the TS in this example). The PV quantity cannot be thus considered, in this case, reliably predicted by the model for the range of time steps considered (and thus for acceptable computational costs).

Other global similarity measures between curves obtained with different values of TS can be also used in addition to the approximation error analysis (Section 3.1). Among them are Root Mean Squared Error (RMSE) and Pearson Correlation Coefficient (PCC). In this study, values of PCC higher than 0.9 were found for Th17, Ab and TotAGS at TS= 10 mins, while with the same time step size a value of PCC of about 0.75 was computed for the most critical output variable B.

The local smoothness analysis reported peak values in correspondence of the two peaks: values up to 890 for output Ab, up to 50 for output Th17, and much lower for the other outputs. For comparison, a sinusoidal signal has D values in the range 5-15, where a random signal shows D values in the range 500-25,000. Although the overall trend of the four time series data can be considered qualitatively smooth, as already reported in Section 2.2, maximum values of D of 50 and 900 would recommend further investigation on the "biological acceptable" concentration rate for the chemical species. The maximum concentration rate predicted by the model for output Ab was 73.7 # entities/ (μL *s), and that for output Th17 was 4.3 # entities/ (μL *s). Experiments reported in the literature [31], [32] confirm these values are plausible.

Results obtained from the so called "parameter sweep" analysis were extremely important, firstly, to ensure that the model behaves reliably for particular input values, secondly, to verify that all the inputs affected the outputs variables with an "expected level" of influence. Thanks to the "once at a time" study, was in fact possible to assess that, for this specific simulated scenario, two inputs (Treg and LXA4) do not have any effect on the outputs results and also that few outputs are completely insensitive to any combination of inputs (FVs of Ab and TotAGS). High values of $p_{BMI,FV\_Th17}$ and $p_{BMI,PV\_Ab}$ suggested that a possible critical effect of BMI on $FV_{Th17}$ and on $PV_{Ab}$ should be further investigated.

### 4.2. Stochastic model verification

The stochastic verification aspects of the model were analysed in the second part of the model verification procedure. In particular, with a fixed value of TS, the model inputs set according to $I_{NIP}$, and varying the $RS_{EF}$ was possible to:

i. identify the "best" fitting shape of the distributions and assess consistency;

ii. define the minimum sample size needed to have a statistical significance.

Symmetry and stability of the output distributions, suggested by the graphical check on the histograms and by the basic statistical summary (Tab. 4), was also confirmed by the results obtained with the Kullback-Leibler divergence analysis. All the outputs seemed to be well approximated by the Gaussian distribution. However, the $r(X)$ was less than one for 4 of the 12 OVs ($TPV_{Th17}$, $FV_{Th17}$, $PV_{TotAGS}$ and $TPV_{TotAGS}$) suggesting that, for these cases, the Student-t is a better fit than the Gaussian.

Regarding the convergence and stability of the mean and standard deviation as the sample size increases, a first graphical check is given by the plots of the cumulative mean and standard deviation (example reported in Fig. 5(a,b)). All of them showed that after a few hundred samples, both the CMean_s and CStd_s converge to a certain value. More formally, the number of runs required to obtained an accurate estimate of the mean and standard deviation was computed assessing the variability of the measurements and evaluating the coefficient of variation rate. The outputs that seemed to converge faster are the $PV_{Ab}$ and $TPV_B$. It is important to consider that, because a noisy behaviour of the quantity *IC* can be observed, some fitting or filtering operations might help to precisely identify *Ns*.

This study has some limitations. A first critical point regards the discretization error analyses in which time is considered the only discrete variable for UISS-TB. It is important to notice that, in most of the cases, also the space domain is usually partitioned in the so called "bins". In addition to the time step convergence analysis, a "bin size" convergence analysis might be also needed.

Regarding the parameter sweep analysis, one assumption made was that all the inputs of the vector of features are independent one from each other which of course this is not generally the case. However, assuming inputs are independent allows to explore all possible combinations, even those that might not make physical / biological sense. This is a conservative assumption.

Another limitation of the study is that the consistency analyses and the sample size determination was made considering only the stochastic variability due the environmental factor. This because in principle the initial distribution of entities and the HLA can be observed in each patient being modelled (the first with a lung radiograph, the second with immune-essays). If these or other inputs cannot be defined precisely, but only in term of population distributions, the sample size could become much larger.

In conclusion, the solution verification of agent-based models is a non-trivial activity. The approach proposed here is exhaustive, in the sense that it covers all possible sources of numerical solution errors. Also, the application to UISS-TB model showed that the proposed approach is effective in highlighting some limitations of the current model implementation, that depending on the context of use might or might not be critical.

### ACKNOWLEDGEMENTS

This study was supported by the STriTuVaD project (SC1-PM-16-2017- 777123). The authors declare that they do not have any financial or personal relationships with other people or organisations that could have inappropriately influenced this study.